------------------------------------------------------------------------------
\documentstyle{article}
\begin{document}
\begin{titlepage}    
\vskip 3mm
\begin{center}
 { \LARGE \bf Heavy Flavor Weak Decays }\\
\vskip 10mm
{R. C. Verma}\\
\vskip 2mm
{\em Department of Physics\\
Punjabi University, Patiala - 147 002, {\bf India}  \\
e-mail: rcv@pbi.ernet.in }\\
\end{center}
\vskip  10mm
\begin{abstract}
Weak decays of heavy flavor hadrons play a special role in our understanding
of physics of the Standard Model and beyond. The measured quantities, 
however, result from a complicated interplay of weak and strong interactions.
Weak leptonic and semileptonic decays are reasonably well understood, 
whereas weak hadronic decays present challenges to theory. In this talk, 
we review the present status of exclusive weak decays of  charm and bottom 
hadrons.
\vskip 30 mm
Talk delivered at XII DAE Symposium on High Energy Physics, Guwahati,
India (Dec. 26, 1996 - Jan. 1, 1997)
\end{abstract}
\end{titlepage}
\newpage
\large
\section{Introduction}
Soon after the discovery of $J/\psi(\bar c c)$ meson in 1974, 
weakly decaying pseudoscalar charm mesons ($D^{0},  D^{+}$ and $D_{s}^{+}$)
were produced. Data on these hadrons have been collected at $e^{+}e^{-}$ 
colliders and at fixed target experiments [1]. 
Study of B-physics began in 1977 with  the discovery of  
$\Upsilon (\bar b b)$ state. However, further progress 
in measurements in naked bottom sector could occur only in the last decade  
with the development of high resolution silicon vertex detector and high 
energy colliders [2,3]. Three 
bottom pseudoscalar mesons ($ B^{+}, B^{0}$ and $B_{s}^{0}$) have been 
studied whereas the fourth meson $B_{c}^{+}$ is also expected to be produced.  
In the baryon sector, a few weakly decaying charm baryons  
($\Lambda_{c}^{+}, \Xi_{c}^{+}, \Xi_{c}^{0}$ and $\Omega_{c}^{0}$) and one 
bottom baryon $\Lambda_{b}$ have been observed experimentally [1]. A number of 
charm and bottom baryons are expected to be seen in future experiments. 

The weak currents in the Standard Model generate 
leptonic, semileptonic and hadronic decays of the heavy flavor hadrons.
An intense activity on theoretical [2-7] and experimental [8-11] studies of 
these hadrons has been going on in this area.
Experimental studies have mainly focused on precision measurements 
of branching ratios for their weak decays. Regarding 
their lifetime patterns, inclusive decays, exclusive leptonic
and semileptonic decays, a complete picture is beginning to emerge [4],
though a few discrepancies yet remain to be explained.
However, a theoretical description of exclusive hadronic decays based 
on the Standard Model is not yet fully possible [3,5] as 
these involve low energy strong interactions. 
Weak decays of heavy quark hadrons provide an ideal 
opportunity to probe strong interactions, to determine the Standard Model
parameters and to search for physics lying beyond the Model.

In this review, present status of exclusive weak decays of heavy flavor
hadrons is given. We first discuss 
their lifetime pattern, leptonic and semileptonic decays.  
Then weak hadronic decays of charm and bottom  
mesons are presented. Particularly, emphasis is given on the 
factorization hypothesis and relating the hadronic modes with the semileptonic 
decays. Finally, baryon decays are briefly introduced. In preparing 
this short talk, it has been difficult to make a complete presentation of 
all the aspects of weak decays. For further information, 
reference is made to some review articles [2-6].
\section { Lifetime Pattern of Heavy Flavor Hadrons}
At quark level various diagrams can contribute to the  weak decays.
These are generally classified as a) Spectator quark, b) W-exchange 
c) W-annihilation and d) Penguin diagrams. 
W-exchange and W-annihilation diagrams are suppressed 
due to the helicity and color considerations. Penguin diagrams, contributing
to Cabibbo suppressed modes, are also expected to be small in strength.
Thus the dominant quark level processes seem to be those in 
which light quark/s behave like spectator. 
This simple picture then immediately yields decay width for a hadron 
containing a b quark, 
$$ \Gamma =  {{ G_{F}^{2} ~ m_{b}^{5} } \over { 192 \pi^{3} }}  
 |V_{bc}|^{2} \times F_{ps},  \eqno(1) $$
where $F_{ps}$ is a phase-space factor.  There is also a term 
with $|V_{bu}|^{2}$, which is very small and has been neglected [2]. Thus all 
the bottom hadrons are predicted to have equal lifetimes. 
For charm hadrons also, the spectator quark model leads to equal 
lifetimes. Though order of estimate
of life-times is alright, the individual values [1] do show deviations 
from a common lifetime:
$$ \tau(D^{+}) \approx 2.5 \tau(D^{0}) \approx 2.5 \tau(D_{s}^{+}) 
\approx 5.0 \tau(\Lambda_{c}^{+}) \approx 3.0 \tau(\Xi_{c}^{+}) \approx 10 
\tau(\Xi_{c}^{0}).  \eqno(2) $$
These differences seem to arise from many  considerations [6], like\\
a) interference among the spectator diagrams (color enhanced \& 
color suppressed) which enhances $D^{+}$ life-time;\\
b) nonspectator diagrams, like W-exchange diagram, which 
yield the following life-time pattern for the charm baryons:
$$ \tau (\Xi_{c}^{0})  < \tau (\Lambda_{c}^{+}) <  \tau (\Xi_{c}^{+}). 
\eqno(3) $$ 
Applying these considerations to the bottom hadrons, following observed 
pattern can be obtained [12]:
$$ \tau( \Lambda_{b}) <  \tau( B^{0}) \approx \tau(B_{s}^{0}) 
\approx \tau(B^{+}).  \eqno(4) $$
However, an exact agreement with experiment for 
B meson and $ \Lambda_{b}$ lifetime ratio is difficult
to obtain. Recently, this ratio is described [13] by a simple ansatz that
 replaces  the quark mass with the decaying hadron mass in the 
$m_{Q}^{5}$ factor in 
front of the hadronic width. However, there is yet no theoretical 
explanation for the ansatz. 
\section {Weak Leptonic and Semileptonic Decays}
In the Standard model, leptonic and semileptonic decays
naturally involve factorizations of their
amplitudes in terms of a well understood leptonic part and a more
complicated hadronic current for the quark transition.
Lorentz invariance is then 
used to express the matrix element in terms of a few formfactors which 
contain the nonperturbative strong interaction effects [4]. Explicit quark 
models [14-20] have been constructed to construct 
the hadron states which are then used to calculate the formfactors.  
In the last few years, a new theoretical approach known as the Heavy Quark
Effective Theory (HQET) has emerged for analyzing heavy flavor hadrons.
In the limit of heavy quarks, new symmetries [21] appear which simplify
the calculations of the formfactors. Nonperturbative approaches like
lattice simulations [22] and QCD sumrules [23] have also been used to 
calculate the formfactors.

Weak quark current generating the charm hadron decays is
$$J_{\mu}^{\Delta C = -1} = V_{cs}^{*} (\bar s c) + V_{cd}^{*} (\bar d c),  
\eqno(5) $$
where $\bar q' q$ denotes the V-A current
$\bar {q}' \gamma_{\mu}(1 - \gamma_{5})q$
and $V_{qq'}$ represents  the corresponding Cabibbo-Kobayashi-Maskawa 
(CKM) matrix element.
Selection rules for these decays are:
$$ \Delta Q = -1, \Delta C = -1, \Delta S = -1~ for ~Cabibbo ~enhanced 
~   c \rightarrow s + l + \nu_{l}~ process,  $$ 
$$  \Delta Q = -1, \Delta C = -1, \Delta S = 0~ for~ Cabibbo~ suppressed~
   c \rightarrow d + l + \nu_{l} ~ process.  $$
Similarly, the weak quark current
$$ J_{\mu}^{\Delta b = 1} = V_{cb} (\bar c b) + V_{ub} (\bar u b),  
\eqno(6) $$
gives the following selection rules for bottom hadron decays:
$$ \Delta Q = 1, \Delta b = 1, \Delta C = 1~ for~ CKM~ enhanced 
 ~  b \rightarrow c + l + \nu_{l} ~process,  $$
$$  \Delta Q = 1, \Delta b = 1, \Delta C = 0 ~for~ CKM~ suppressed
 ~  b \rightarrow u + l + \nu_{l} ~process.  $$  
\subsection {Leptonic Decays: $P(J^{P}=0^{-}) \rightarrow l +\nu_{l} $}
These decays are the simplest to consider theoretically, and are usually 
helicity suppressed particularly when lighter leptons are emitted [24, 25]. 
Decay amplitude for a typical decay $ D \rightarrow l \nu_{l}$ 
involves the decay constant $f_{D}$ defined as
$$ <0|A_{\mu}|D(p) > = i f_{D} p_{\mu}. \eqno(7) $$
which measures the amplitude for the quarks to have zero separation.
This leads to the following decay width formula:
$$\Gamma (D(\bar q c) \rightarrow l \nu_{l}) = {G_{F}^2 \over {8 \pi}}
|V_{q c}|^{2} f_{D}^{2} m_{D} m_{l}^{2} (1 -  {m_{l}^{2} 
\over m_{D}^{2}})^{2}. \eqno(8) $$
For $D^{+} \rightarrow l \nu_{l}$ decay, all the theoretical values [4] for 
$f_{D}$, ranging from 170 MeV to 240 MeV, are well below the experimental 
limit [26]:
$$f_{D} < 310 MeV. \eqno(9) $$
For $ D_{s}^{+} \rightarrow l \nu_{l}$ decay,
Particle Data Group [1, 27] gives the following values:
$$f_{D_{s}} = 232 \pm 45 \pm 20 \pm 48 MeV, ~ 
 344 \pm 37 \pm 52 \pm 42 MeV,~
430 ^{+150}_{-130} \pm 40 MeV. \eqno(10) $$
using the Mark and CLEO data. Potential models [4] give $f_{D_{s}}$ between
190 MeV and 290 MeV. Lattice calculations [22, 28] yield: 
$f_{D_{s}} = 220 \pm 35 MeV$. 
which matches with QCD sumrules estimates[23].
More recently, E653 collaboration [29] has obtained 
$f_{D_{s}} = 194 \pm 35 \pm 20 \pm 14 MeV$ and 
CLEO result has been updated to [30]:
$f_{D_{s}} = 284 \pm 30 \pm 30 \pm 10 MeV. $

For B -mesons, leptonic decays are strongly suppressed by the small value of
$|V_{ub}|^{2}$. Lattice simulations give $f_{B} = 180 \pm 40 MeV$
whereas the scaling law derived in HQET [21],  
$$f_{P} = { { A} \over { \sqrt { M_{P} } } }  
[ \alpha_{s}(M_{P})^{-2/\beta_{0}} \times (1 + O(\alpha) +...]  \eqno(11) $$
predicts a rather lower estimate $f_{B} = 120 MeV$ [28] which is expected
to increase due to the radiative corrections. 
Potential model values [4] 
range from 125 MeV to 230 MeV. QCD sumrules estimate:
$f_{B} = 180 \pm 50 ~ MeV$ is in good agreement with those from the lattice
calculations. Thus, theory predicts [3]
$$ B(B^{+} \rightarrow \tau^{+} \nu_{\tau} ) \approx 4.0 \times 10^{-5}, $$
for the most accessible of the leptonic B decays because the large $\tau$ mass
reduces the helicity suppression.
Experimentally, the following upper limit is available [1]:
$$B(B^{+} \rightarrow \tau^{+} \nu_{\tau} ) < 1.8 \times 10^{-3}. $$

Measurement of $f_{B}$ decay constant at future b-factories would have a 
significant impact on the phenomenology of heavy flavor decays.
A precise knowledge of $f_{B}$ would allow an 
accurate extraction of the  CKM matrix element $|V_{ub}|$. Moreover, 
it enters into many other B-decay measurements, notably $B-\bar B$ mixing
and CP violation in B-decays  [3, 31, 32].
The standard model allows $B, B_{s} \rightarrow l^{+} l^{-} $ leptonic decays
via box or loop diagrams. Theoretical values [3, 33] are well below the present 
experimental limits [1]. 
\subsection {Semileptonic Decays: $ P \rightarrow M(J^{P}=0^{-}~ or~1^{-})
 + l + \nu_{l} $ }
With the enormous data samples now available for charm and bottom mesons, 
their semileptonic decays, particularly emitting a pseudoscalar 
meson or a vector meson, are well measured.  These decays occur 
via spectator quark diagram and involve no final state interactions. 
So these decays are the primary source of the CKM elements and various 
formfactors.
Decay amplitude for $ P(\bar q' Q) \rightarrow M(\bar q' q) l \nu_{l}$ 
is given by 
$$ <M(\bar q' q) l \nu_{l}|H_{W}^{semi-lep}|P(\bar q' Q)> = 
{ {G_{F}} \over {\sqrt {2}} } V_{qQ}^{*} <M |J_{\mu}(\bar q Q)|P> (\nu_{l}
\gamma^{\mu} (1 - \gamma_{5})l).  \eqno(12) $$ 
Using Lorentz invariance, the hadronic matrix elements are described by a few 
formfactors which are also needed in the analysis of the weak hadronic decays.
\subsubsection {Semileptonic Decays of Charm Mesons}
{\bf $D \rightarrow  P l \nu_{l} $ Decays }\\
When the final state meson is pseudoscalar, parity implies that 
only the vector component of the weak current contributes to the decay,
whose matrix element is given by [6, 14], 
$$<P(p')|V_{\mu}| D(p) > =  [ (p+p')_{\mu} - { {m_{D}^{2} - m_{P}^{2}} 
\over {q^{2}}} 
 q_{\mu} ] F_{1}(q^{2}) + { {m_{D}^{2} - m_{P}^{2}} \over {q^{2}}} 
 q_{\mu} F_{0}(q^{2}), \eqno(13) $$
where $F_{1}(0)=F_{0}(0)$ and $q_{\mu} = (p-p')_{\mu}$. The formfactors 
represent the amplitude that the final state $(\bar q s)$ pair forms a K meson. 
Energy of K meson in the rest frame of D meson is linearly related to $q^{2}$,
$$E_{K} = { { m_{D}^{2}+m_{K}^{2}-q^{2}} \over {2 m_{D}} }. \eqno(14) $$
At $q^{2} = q_{max}^{2} = (m_{D}-m_{K})^{2}$, the K meson is at rest in the 
rest frame of D meson. Then the overlap of the initial and final state is 
maximum and so the formfactor is at its maximum value. 
At $q^{2} = 0, E_{K}$ is maximum and so the formfactor is at its minimum value. 
This $ q^{2} $ dependence is usually
expressed through the pole dominance formula [14],
$$F(q^{2}) = { { F(0)} \over { 1- {q^{2}/m_{*}^{2}} } }, \eqno(15) $$ 
which is studied by measuring the differential decay rate [4]. 
Present data [4, 34] on differential decay rate for 
$D \rightarrow \bar K l \nu_{l}$ yields, for 
$|V_{cs}| = 0.974$ and the pole mass $ m^{*} = 2.00 \pm 0.11 \pm 0.16 GeV.,$
$$ F_{1}^{DK}(0) = 0.75 \pm 0.03.  \eqno(16) $$ 
Quark model values lie between 0.7 to 0.8 [14-20], lattice calculations give
0.6 to 0.9 [22] and QCD sumrules approach gives 0.6 [23] for this formfactor.  

Decay width ratio of Cabibbo suppressed decay 
$D \rightarrow \pi l \nu_{l}$ and  the $D \rightarrow \bar K l \nu_{l}$
serves to deliver $F_{1}^{D \pi}/F_{1}^{D\bar K}$. Mark III and CLEO
data [34] yield the following respective values:
$$F_{1}^{D \pi}/F_{1}^{D\bar K} = 1.0_{-0.3}^{+0.6} \pm 0.1, ~~
 1.29 \pm 0.21 \pm 0.1.  \eqno(17) $$
These results are consistent with theoretical predictions which range
from 0.7 to 1.4 [4].

{\bf {$D \rightarrow V(J^{P}=1^{-}) l \nu_{l}$ } Decays}\\
When the final state meson is a vector meson, there are four independent 
form factors [14]:
$$<V(p', \epsilon )|J_{\mu}| D(p) > = {{ 2} \over {m_{D} + m_{V}} }
\epsilon_{\mu \nu \rho \sigma} \epsilon^{* \nu} p^{\rho} p'^{\sigma} V(q^{2})
+i  [ (m_{D} +m_{V}) \epsilon_{\mu}^{*} A_{1}(q^{2}) $$
$$  - { {\epsilon^{*} \cdot q } \over { m_{D} + m_{V}}} (p +p')_{\mu} 
A_{2}(q^{2}) - 2 m_{V} { {\epsilon^{*} \cdot q } \over { q^{2}} } q_{\mu} 
(A_{3}(q^{2}) - A_{0}(q^{2}))], \eqno(18) $$
where $\epsilon_{\mu}$ is the polarization vector of the vector meson, and 
$q_{\mu} = (p - p')_{\mu}$ is the momentum transfer.
Total decay width $\Gamma(D \rightarrow \bar K^{*} l \nu_{l})$ is dominated by 
$A_{1}$ formfactor. Ratios of other formfactors $V$ and $A_{2}$ with $A_{1}$ 
are determined from the angular distribution [2-4]. Present data [34] yield:
$$ A_{1}^{D \bar K^{*}}(0) = 0.56 \pm 0.04,
A_{2}^{D \bar K^{*}}(0) = 0.40 \pm 0.08, 
V^{D \bar K^{*}}(0) = 1.1 \pm 0.2.  \eqno(19) $$
Theoretically quark models [14-20] give large values   
$ A_{1} (0)$ = 0.80 to 0.88 and $A_{2} (0)$ = 0.6 to 1.2, whereas the
predictions for $V(0)$ range from 0.8 to 1.3 in good agreement with 
experiment.
Lattice calculations [22] and QCD sumrules estimates [23] are in better
agreement with experiment [4].

For Cabibbo suppressed mode, experimental value [1]
$$ B(D^{+} \rightarrow  \rho^{0} \mu^{+} \nu_{\mu}) /
 B(D^{+} \rightarrow  \bar K^{*0} \mu^{+} \nu_{\mu}) =
0.044 _{-0.025}^{+0.031} \pm 0.014,  \eqno(20) $$
is consistent with theoretical expectations [4, 18] with in the errors.

Semileptonic decays of strange-charm meson ($D_{s} \rightarrow \phi /
\eta /\eta' + l +\nu_{l}$) have also been measured [1]. These decays appear
to follow the pattern of D decays in terms of the formfactor ratios [4].
\subsubsection {Semileptonic Decays of B Mesons}
For B-decays, following data is available for CKM enhanced mode [1]:
$$B(B^{0} \rightarrow D^{-} l^{+} \nu) = 1.9 \pm 0.5 \%,  $$
$$B(B^{0} \rightarrow D^{*-} l^{+} \nu) = 4.56 \pm 0.27 \%,  $$
$$B(B^{+} \rightarrow \bar D^{0} l^{+} \nu) = 1.6 \pm 0.7 \%, $$
$$B(B^{+} \rightarrow \bar D^{*0} l^{+} \nu) = 5.3 \pm 0.8 \%. $$
Using $|V_{cb}| = 0.038 \pm 0.004$, present data yield [34, 35]
$$ A_{1}(0) = 0.65 \pm 0.09, $$
$$ V(0)/A_{1}(0) = 1.30 \pm 0.36 \pm 0.14, ~~
 A_{2}(0)/A_{1}(0) = 0.64 \pm 0.26 \pm 0.12, \eqno(21) $$
which are consistent with quark models estimates [4].

In nonperturbative problems, exploitation of all the available symmetries 
is very important. For the heavy flavor physics, the use of spin-flavor 
symmetries that are present when masses of the heavy quarks are
 $>> \Lambda_{Q}$, leads to considerable simplifications [21].
In going to the limit $m_{c}, m_{b} \rightarrow \infty$, all the formfactors
are expressed in terms of one universal function called 
Isgur-Wise function $ \zeta (\omega)$, 
$$ F_{1}(q^{2}) =  V(q^{2}) =  A_{0}(q^{2}) =  A_{2}(q^{2})   $$
$$ =  [ 1- { {q^{2} } \over { (M_{B}+M_{D})^{2} } } ] ^{-1} A_{1} (q^{2}) 
=  { {M_{B} + M_{D} } \over  { 2 \sqrt { M_{B} M_{D} } } } \zeta (\omega ), 
  \eqno(22) $$
where  $ \omega = v_{B} \cdot v_{D}$. These relations are valid 
up to perturbative and power corrections [4, 28]. Theoretical difficulty in 
making predictions for the form factors lies in calculating these 
corrections with sufficient precision.
At present, in the presence of these corrections, 1.30 and 0.79  
are obtained [4, 35] for the ratios $V/A_{1}$ and $A_{2}/A_{1}$ respectively.

Charmless semileptonic branching fraction is expected 
to be around $1\%$ of that of the semileptonic decays emitting charm meson
based on the present estimate $|V_{ub}/V_{cb}| = 0.08 \pm 0.02$ [1]. 
Heavy quark symmetry is less predictive for heavy $\rightarrow $ light decays
 than it is for heavy $\rightarrow $ heavy ones. 
Experimentally two branching ratios have been measured recently by CLEO 
collaboration [36]:
$$ B(B^{0} \rightarrow \pi^{-} l^{+} \nu) = (1.8 \pm 0.4 \pm 0.3 \pm 0.2
)\times 10^{-4}, $$
$$B(B^{0} \rightarrow \rho^{-} l^{+} \nu) = (2.5 \pm 0.4^{+0.5}_{-0.7}
 \pm 0.5) \times 10^{-4}, $$
which are consistent with theoretical expectations.

In addition to single meson emission in the 
final state, semileptonic decays also permit the production of two or more
mesons. Quite often these mesons are produced through decay of a meson 
resonance produced in the weak decays [1]. For D-mesons, known resonant 
exclusive modes come close to saturating the inclusive semileptonic rates. 
In B decays, there is some room for nonresonant multi-hadron final state. 
Semileptonic decays of charm and 
bottom baryons have also been observed. 
However, experimental results currently have limited statistical significance.
Much larger data on these decays are expected in the future, allowing tests of 
various theoretical models [37].
\section {Weak Hadronic Decays }
Weak hadronic decays of heavy flavor hadrons are considerably 
complicated to treat theoretically. At current level
of understanding these require model assumptions. Even if the short distance 
effects due to hard gluon exchange can be resummed and the effective 
Hamiltonian has been constructed at next to leading order, evaluation 
of its matrix elements is not straightforward. 
Various theoretical and phenomenological approaches have been
employed to study weak hadronic decays. Broadly, these are:\\
\vskip 1mm 
{\bf i) Flavor Symmetry Frameworks}\\
In the flavor symmetry frameworks, initial and final state mesons 
and weak Hamiltonian belong to their irreducible representations. Using 
Wigner-Eckart theorem, decay amplitudes are expressed in terms of few reduced
amplitudes. Thus useful sumrules among different decay amplitudes are
obtained [38]  using isospin and SU(3) flavor symmetries. However, SU(3)
violation has been shown by the charm meson decay data [39]. \\
\vskip 1mm 
{\bf ii) Quark Line Diagram Approach}\\
Quark diagrams appearing in the weak decays are classified according 
to the topology of weak 
interaction with all the strong interaction effects included. With each 
quark line diagram, a corresponding parameter is attached and appropriate 
C.G. coefficients are introduced depending upon the initial and final state
particles [40]. Using experimental values, relative strengths of various 
quark diagrams are then obtained.\\
\vskip 1mm 
{\bf iii) Relativistic and Nonrelativistic Quark Models}\\
Explicit quark model calculations have been done to determine the 
strength of various quark level processes. These models usually employ 
factorization [41] which can be used to relate hadronic decays to the 
semileptonic decays [42].\\
\vskip 1mm 
{\bf iv) Nonperturbative Methods}\\
QCD sumrules [23] approach has provided the general trends but agreement 
with present data is poor at a quantitative level. 
Lattice QCD calculations [22], though promising, are still
in progress. Further these methods have their own uncertainties.\\

At present extensive data [1, 43] exist for weak hadronic decays of 
charm and bottom mesons; though in the baryon sector, only a few decay modes 
of $\Lambda_{c}^{+}, \Xi_{c}^{+}$ and $\Lambda_{b}$
have been studied experimentally [1, 44]. 
The heavy flavor hadrons have many channels available for their decay 
involving two or more hadrons in their final states. However, for charm hadron
decays, two-body decays dominate the data as multibody decays
show resonant structure. Due to the considerably larger phase space that 
is available in bottom hadron decays and to the much higher number of open
channels such a feature cannot extend to the bottom hadrons. Nevertheless
these are expected to make up significant fraction of their hadronic decays.

Most of the observed two-body decays of heavy flavor mesons involve 
pseudoscalar (P) and vector (V) mesons (s-wave mesons) in their final state:  
$ P \rightarrow PP/PV/VV.$ 
In addition, some of the decays of charm mesons
emitting axial (A), Scalar (S) and tensor (T) mesons (p-wave mesons), like
$ P \rightarrow P + A/S/T $
have also been measured [1]. Bottom mesons, being massive, can also decay to 
vector meson and another p-wave meson, or two p-wave mesons. 
In addition to these modes, weak decays accompanying photon 
(like $ B \rightarrow K^{*} + \gamma$) are also observed.
\subsection {Weak Hadronic Decays of Charm  Mesons}
The general weak current $\otimes$ current weak Hamiltonian for hadronic  
weak decays in terms of the quark fields is given by  
$$ H_{W}^{ \Delta C= \Delta S = -1 }=
 {G_{F}\over {\sqrt 2}} V_{ud}V_{cs}^{*} (\bar u d)( \bar s c),
\eqno(23a)$$
for Cabibbo enhanced mode,
$$ H_{W}^{\Delta C= -1, \Delta S = 0} =
 {G_{F}\over {\sqrt 2}}  [V_{us}V_{cs}^{*} (\bar u s)( \bar s c) +
 V_{ud}V_{cd}^{*} (\bar u d)( \bar d c)],
\eqno(23b) $$
for Cabibbo suppressed  mode, and  
$$ H_{W}^{\Delta C= - \Delta S = -1 } =
 {G_{F}\over {\sqrt 2}} V_{us}V_{cd}^{*} (\bar u s)( \bar d c),
\eqno(23c) $$
for doubly Cabibbo suppressed  mode.
Since only quark fields appear in the weak Hamiltonian, the weak hadronic 
decays are seriously affected by the strong interactions. 
One usually identifies the two scales [6] in these decays: 
short distance scale at 
which W-exchange takes place and long distance scale where final state 
hadrons are formed.  As the hard gluon effects 
at short distances are calculable using the perturbative QCD,  
long distance effects, being nonperturbative, are the 
source of major problems in understanding the weak hadronic decays. 
The hard gluon exchanges renormalize the weak vertex and introduce new 
color structure [6].
Effective weak Hamiltonian thus acquires effective neutral current term. 
For instance, weak Hamiltonian for Cabibbo enhanced mode becomes
$$ H_{W}^ { \Delta C= \Delta S = -1 }=
 {G_{F}\over {\sqrt 2}} V_{ud}V_{cs} [ c_{1} (\bar u d) (\bar s c)
+ c_{2} (\bar s d) (\bar u c)], \eqno(24) $$
where the QCD coefficients $ c_{1} = {1 \over 2}(c_{+} + c_{-})$,
$c_{2} = {1 \over 2}(c_{+} - c_{-})$ and $c_{\pm}(\mu)=  [\frac 
{\alpha_{s}(\mu^{2})} {\alpha_{s}(m_{W}^{2})}]^{d_{\pm}/2b}$ with 
$d_{-}= -2d_{+}=8$ and $b= 11- {2\over 3}N_{f}$,
$N_{f}$ being the number of effective flavors, $\mu$ the mass scale [6]. 
\subsubsection { $ D \rightarrow PP/PV/VV $ Decays}
Decay width for a two-body decay of D meson is given by 
$$ \Gamma (D \rightarrow M_{1} + M_{2}) = G_{F}^{2} (CKM~factors)^{2} 
k^{2l+1} $$
$$ \times  \sum_{i} (mass ~ factors) |< M_{1} M_{2})| O_{i} |D>|^{2},  
\eqno(25) $$
where $ l $ denotes the angular momentum between the final state mesons 
$M_{1}, M_{2}$, and $ i $ denotes the helicity of these mesons. The 
operators $ O_{i} $ correspond to the quark processes responsible for 
the decays. In the evaluation of matrix element of the 
weak Hamiltonian, one usually applies the factorization hypothesis
 [6, 14] which expresses hadronic decay amplitude as
the product of matrix elements of weak currents between meson states, 
$$ <P_{1} P_{2} | H_{w}| D> ~\propto~ <P_{1}| J_{\mu} |0>
<P_{2}|J^{ \dagger \mu} |D>,  \eqno(26a) $$
$$ <P V | H_{w}| D>~ \propto~[
<P |J_{\mu} |0><V|J^{ \dagger \mu} |D> ~+~ <V| J_{\mu} |0><P|J^{ \dagger \mu}
 |D>],  \eqno(26b) $$
$$ <V_{1} V_{2} | H_{w}| D> ~ \propto ~ 
<V_{1}| J_{\mu}|0><V_{2}|J^{ \dagger \mu } |D>.  \eqno(26c) $$
Matrix elements of weak current between meson and vacuum state are given by
eq.(7) and  
$$<V(p, \epsilon )|J_{\mu}| 0> = \epsilon^{*}_{\mu} m_{V} f_{V}. \eqno(27) $$
Meson to meson matrix elements appearing here have already been given in eqs.
(13) and (18). Thus factorization scheme allows us to
predict decay amplitudes of hadronic modes in terms of the semileptonic 
formfactors and meson decay constants. 

For the sake of illustration, we consider Cabibbo enhanced decays 
$D \rightarrow P P $. Separating the factorizable and nonfactorizable parts, 
the matrix element  of the weak Hamiltonian, given in eq. (24), 
between initial and final states can be written [6, 45] as
$$ < P_{1} P_{2} | H_{w} | D > ~ = ~
{ G_{F} \over {\sqrt 2} } V_{ud}V_{cs}^{*} 
 [ a_{1} < P_{1}| (\bar u d)|0>< P_{2}|(\bar s c) | D >  $$
$$ +  a_{2} < P_{2}| (\bar s d)|0>< P_{1}|(\bar u c) | D >   $$
$$  + (c_{2}   < P_{1} P_{2} |  H_{w}^{8} | D >  
  +  c_{1} < P_{1} P_{2} | \tilde{H}_{w}^{8} | D >)_{non ~fac} ] \eqno(28) $$
where
$$ a_{1,2} = c_{1,2} + \frac {c_{2,1}} {N_{c}},  $$
$$ H^{8}_{w} ~~ =
~~ \frac {1} {2} \sum_{a=1}^{8}  ( \bar u \lambda ^{a} d ) (
\bar s \lambda ^{a} c ), \hskip 0.2 cm
\tilde{H} ^ {8}_ {w} ~~ = ~~ \frac {1} {2} \sum_{a=1}^{8}
( \bar s \lambda ^{a} d ) ( \bar u\lambda ^{a} c ).    $$

In addition, nonfactorizable effects may also arise through 
the color singlet currents [46]. Matrix elements of the first and the second
terms in eq. (28) can be calculated using the factorization scheme. 
So long as one restricts to the color singlet
intermediate states, remaining terms are usually ignored and one  
treats $a_{1}$ and $a_{2}$ as input parameters in place of using $N_{c}
= 3$ in reality. 
The charm hadron decays are classified in three classes, namely\\
 i) Class I transitions that depend solely on $a_{1}$ (color favored),\\
 ii) Class II transitions that depend solely on $a_{2}$ (color suppressed),\\
 iii) Class III transitions that involve interference between terms with $a_{1}$ and $a_{2}$.\\
It has been believed  [6, 14] that the  
charm meson decays favor $N_{c} \rightarrow \infty $ limit, i.e.,
$a_{1} \approx 1.26, \hskip 0.5cm a_{2} \approx -0.51.$, indicating
destructive interference in $D^{+}$ decays.
\subsubsection {Long Distance Strong Interaction Effects}
The simple picture of spectator quark model works well in 
giving reasonable estimates for the exclusive
semileptonic decays. However, success in predicting 
individual hadronic decays is rather limited. For example, spectator quark 
model yields the following ratios:
$$ { { \Gamma (D^{0} \rightarrow \bar K^{0} \pi^{0}) } \over {\Gamma (D^{0} 
\rightarrow K^{-} \pi^{+})  }} = 0.1~~ (0.5 \pm 0.1~~Expt.)  \eqno(29a) $$
for Cabibbo enhanced mode and 
$$ { { \Gamma (D^{0} \rightarrow K^{+} K^{-}) } \over {\Gamma (D^{0} 
\rightarrow  \pi^{+} \pi^{-}) }} = 0.9~~ (2.5 \pm 0.4~~ Expt.)  \eqno(29b) $$
for Cabibbo suppressed mode.

Similar problems exist for $ D \rightarrow \bar K^{*} \pi / \bar K^{*} 
\rho $ decay widths. Besides these, 
other measured decays, involving  single isospin
final state, also show discrepancy with theory.
 For instance, the observed $D^{0}
\rightarrow \bar {K^{0}} \eta $ and $ D^{0} \rightarrow \bar {K^{0}}
\eta'$ decay widths are considerably larger than those predicted in the
spectator quark model. Also measured
branching ratios for $ D^{0} \rightarrow \bar {K^{*0}} \eta $, $
D_{s}^{+} \rightarrow \eta / \eta' + \rho^{+},$ are found to be higher than 
those predicted by the spectator quark diagrams. 
For $ D_{s}^{+} \rightarrow
\eta / \eta' + \pi^{+},$ though factorization can account for
substantial part of the measured branching ratios, it fails to relate
them to corresponding semileptonic decays $ D_{s}^{+} \rightarrow \eta /
\eta' + e^{+} \nu $ consistently [47]. 

In addition to the spectator quark
diagram, factorizable W-exchange or W-annihilation diagrams may contribute
to the weak nonleptonic decays of D mesons. However, for $ D \rightarrow PP $
decays, such contributions are helicity suppressed. 
For $ D $ meson decays,
these are further color-suppressed as these involve QCD coefficient $c_{2}$, 
whereas for $ D_{s}^{+} \rightarrow PP $ decays these vanish due to the 
conserved vector (CVC) nature of the isovector current $( \bar u d)$ [47].

It is now established that the factorization scheme does not work well for 
the charm meson decays.
The discrepancies between theory and experiment are attributed to various 
long distance effects which are briefly discussed in the following.\\
\vskip 1mm 
{\bf i) Final State Interaction Effects}\\
Elastic final state interactions (FSI) 
introduce phase shifts in the decay amplitudes [48], which can  be analyzed  
in the isospin framework.
For instance, the 
isospin amplitudes 1/2 and 3/2 appearing in $ D \rightarrow \bar K \pi$
decays may develop different phases leading to;
$$ A(D^{0} \rightarrow K^{-} \pi^{+} )
~~ = ~~ { {1}  \over { \sqrt {3} } }  [ A_{3/2} e^{i
\delta_{3/2}} + \sqrt{2} A_{1/2} e^{i \delta_{1/2}} ], \eqno(30a) $$
$$ A(D^{0} \rightarrow \bar K^{0} \pi^{0} ) ~~ = ~~{ {1 } \over 
 { \sqrt{3} } }  [ \sqrt{2} A_{3/2} e^{i \delta_{3/2}} - A_{1/2} e^{i
\delta_{1/2} } ],  \eqno(30b) $$
$$ A(D^{+} \rightarrow \bar K^{0} \pi^{+} ) \hskip
0.5 cm = \hskip 0.5 cm \sqrt{3} A_{3/2} e^{ i \delta_{3/2} }. \eqno(30c) $$
Similar treatment can be performed for $ D \rightarrow \bar K^{*} \pi,
\bar K \rho, \bar K^{*} \rho$ modes. 
 These decays are seriously affected 
as their final states lie close to meson resonances. 
Experimental data on  these modes yield [48, 49]:
$$ |A_{1/2}|/A_{3/2}| = 3.99 \pm 0.25 ~ and ~ \delta_{3/2}-\delta_{1/2}
 = 86 \pm 8^{o} ~ for ~\bar K \pi ~mode,  $$
$$ |A_{1/2}|/A_{3/2}| = 5.14 \pm 0.54 ~ and ~ \delta_{3/2}-\delta_{1/2}
 = 101 \pm 14^{o} ~ for~ \bar K^{*} \pi~ mode,  $$
$$ |A_{1/2}|/A_{3/2}| = 3.51 \pm 0.75 ~ and ~ \delta_{3/2}-\delta_{1/2}
 = 0 \pm 40^{o} ~ for~ \bar K \rho~ mode, $$
$$ |A_{1/2}|/A_{3/2}| = 5.13 \pm 1.97 ~ and ~ \delta_{3/2}-\delta_{1/2}
 = 42 \pm 48^{o} ~ for~ \bar K^{*} \rho ~mode $$
for Cabibbo enhanced mode, and
$$ |A_{0}|/A_{2}| = 3.51 \pm 0.75 ~ and ~ \delta_{0}-\delta_{2}
 = 0 \pm 40^{o} ~ for~ \pi \pi~ mode, $$
$$ |A_{0}|/A_{1}| = 3.51 \pm 0.75 ~ and ~ \delta_{0}-\delta_{1}
 = 0 \pm 40^{o} ~ for~ \bar K K~ mode, $$
for Cabibbo suppressed mode. 

 In addition to the elastic scattering, inelastic FSI can couple 
different decay channels. For example,
$D \rightarrow \bar K^{*} \pi$ and $D \rightarrow 
\bar K \rho $ decays are found to be affected by such inelastic FSI [48].\\
\vskip 1mm 
{\bf ii) Smearing Effects} \\
Further, in certain decays a wide resonance is emitted, like $ D \rightarrow 
\bar K \rho $. The large width of the meson modifies the kinematical phase 
space available to the decay. These effects can be studied using a running
mass $(m)$ of the resonance, and then averaging is done by introducing an 
appropriate measure $r(m^{2})$ like Breit-Wigner formula. For instance, 
$D \rightarrow P \rho$ decay width is calculated as [50]:
$$\bar \Gamma( D \rightarrow P \rho) = 
\int_{2 m_{\pi}}^{m_{D}-m_{P}} r(m^{2}) \Gamma (D\rightarrow P \rho(m^{2}))
dm^{2}.  \eqno(31) $$
Such effects can be as large as 25 \%. For example, 
$$\bar \Gamma(D^{0} \rightarrow K^{-} \rho^{+}) /
 \Gamma(D^{0} \rightarrow K^{-} \rho^{+}) = 0.77. \eqno(32)   $$
Smearing effects have been studied [51] for $D\rightarrow VV $ also.\\
\vskip 1mm 
{\bf iii) Nonfactorizable Contributions} \\
Indeed factorization, combined with the
assumption that FSI are dominated by nearby resonances, has been in use for
the description of charm meson decays. Recently, this issue has been reopened. 
In the factorization scheme, one works in the large $N_{c}$
limit,  and ignores the nonfactorizable terms, which behave like $1/N{c}$.  
However, this approach has failed when extended to B meson decays [52]. 
So D-meson decays
are being reanalyzed keeping the `canonical' value $N_{c} = 3$, 
real number of colors. 
Efforts have been made to investigate the nonfactorizable contributions.
It is well known that nonfactorizable terms cannot be determined
unambiguously without making some assumptions [45] as these involve
nonperturbative effects arising due to the soft-gluon exchange. 
QCD sumrules approach has been used to estimate them [53], 
but so far these have not
given reliable results. In the absence of exact dynamical calculations, 
search for a systematics in the required nonfactorizable contributions 
has been made using isospin [54] and SU(3)-flavor-symmetries [46]. 
\subsubsection {$D \rightarrow P(0^{-})~ +~ p-wave~ Meson 
~(0^{+}, 1^{+}, 2^{+})$ Decays}
Weak hadronic decays involving mesons of intrinsic orbital momentum $ l > 0$ 
in final state are expected to be kinematically suppressed.  
Some measurements are available on these decays. 
Contrary to the naive expectations, 
their branching are found to be rather large [1]. Estimate for 
formfactors appearing in the matrix elements $<p-wave~ meson~ | J | D >$
are available only in the nonrelativistic ISGW quark model [17, 18]. 
In general, theoretical values are lower than the experimental ones [55].
\subsection { Weak Hadronic Decays of B-mesons}
Weak Hamiltonian involving the dominant $ b\rightarrow c$ transition [2, 3] is 
given by
$$H_{W}^{\Delta b = 1} = {{G_{F}} \over {\sqrt{2}} }  [ 
V_{cb}V_{ud}^{*}(\bar c b)(\bar d u) + 
V_{cb}V_{us}^{*}(\bar c b)(\bar s u)  $$
$$ + V_{cb}V_{cd}^{*}(\bar c b)(\bar d c) + 
V_{cb}V_{cs}^{*}(\bar c b)(\bar s c) ]. \eqno(33) $$ 
A similar expression can be obtained for decays involving
$b \rightarrow u$ transition by replacing $\bar c b$ with $\bar u b$.
Following $\Delta b = 1$ decays modes are allowed:\\
i) CKM enhanced modes:
$$  \Delta C = 1, \Delta S = 0, ~~and~~ 
 \Delta C = 0, \Delta S = -1; \eqno(34a) $$ 
ii) CKM suppressed modes:
$$ \Delta C = 1, \Delta S = -1, ~~and~~
\Delta C = 0, \Delta S = 0; \eqno(34b) $$  
iii) CKM doubly suppressed modes:
$$ \Delta C = -1, \Delta S = -1, 
~and~ \Delta C = -1, \Delta S = 0. \eqno(34c) $$
These provide a large number of decay products to B-hadrons. 
Including hard gluon exchanges, the effective 
Hamiltonian can be written as
$$H_{eff} = { { G_{F}} \over {\sqrt{2}  } } V_{cb} V_{ud}^{*} \{
   a_{1} [ (\bar d u) (\bar c b) + (\bar s c) (\bar c b) ] + 
   a_{2} [ (\bar c u) (\bar d b) + (\bar c c) (\bar s b) ] \}.  \eqno(35) $$
In the large $N_{c}$ limit, one would expect:
$$a_{1} \approx c_{1} = 1.1,  ~~ a_{2} \approx c_{2} = -0.24. $$ 
\subsubsection { Determination of $a_{1}$ and $a_{2}$}
Like charm meson decays, depending upon the quark content of mesons involved, 
B-meson decays can also be classified in the three categories.
Several groups have developed models of hadronic B-decays based on the
factorization hypothesis [2, 3]. Recently, it has also been argued 
that the factorization hypothesis is expected to hold better in the heavy 
quark limit [56], for some decay channels, as the ultra-relativistic final 
state mesons don't have time to exchange gluons.
Present data seem to go well with theoretical expectations for most of the
B-meson decays [3]. For instance,
$${ { B(\bar B^{0} \rightarrow D^{*+} \rho^{-} )} \over {B(\bar B^{0} 
\rightarrow D^{*+} \pi^{-} ) }} =  2.8~ 
(2.59 \pm 0.67 ~~Expt.), \eqno(36a) $$
$${ { B(\bar B^{0} \rightarrow D^{*+} a_{1}^{-} )} \over {
B(\bar B^{0} \rightarrow D^{*+} \pi^{-} ) }} = 3.4 
(4.5 \pm 1.2 ~~Expt.). \eqno(36b)$$
By comparing $B^{-}$ and $\bar B^{0}$ decays, $|a_{1}|, |a_{2}|$ and
the relative sign of $a_{2}/a_{1}$ can be determined. 
Thus $\bar B^{0} \rightarrow 
D^{+} \pi^{-}/D^{+} \rho^{-}/D^{*+} \pi^{-}/D^{*+} \rho^{-}$ yield:
$$|a_{1}| = 1.03 \pm 0.04 \pm 0.16, \eqno(37a) $$
$\bar B^{0} \rightarrow \psi X$ decays yield:
$$|a_{2}| = 0.23 \pm 0.01 \pm 0.01, \eqno(37b) $$
and data on $B^{-} \rightarrow 
D^{0} \pi^{-}/D^{0} \rho^{-}/D^{*0} \pi^{-}/D^{*0} \rho^{-}$ 
clearly yield [3, 52]:
$$a_{2}/a_{1} = 0.26 \pm 0.05 \pm 0.09. \eqno(37c) $$
Note that though magnitude of the ratio is in agreement with theoretical
expectation, its sign is opposite. indicating constructive interference 
in $B^{-}$ decays. Other uncertainties of decay constants, 
FSI and formfactors may change its value but not its sign [3].
 This situation is 
in contrast to that in the charm meson decays, where the ratio $a_{2}/a_{1} 
= -0.40$ implies destructive interference in $D^{+}$ decays. 
Interestingly, the constructive interference enhances the hadronic decay 
width of $B_{u}$ meson and reduce its semileptonic branching ratio [57]
bringing it closer to the experimental value.
\subsubsection {Final State Interaction}
Factorization breaks down in the charm sector due to the presence of
final state interactions. The strength of such long distance effects 
in B-decays can also be determined by performing the isospin analysis of 
related channels, such as $B\rightarrow D \pi$ decays.
At present level of experimental precision, there is no evidence for nonzero
isospin phase shifts in B-decays, as the data gives [3] 
$ cos(\delta_{1/2} - \delta_{3/2}) > 0.82 ~~ for ~~B \rightarrow D \pi. $
\subsubsection { Tests of Factorization}
Since a common matrix element $<M | J | B> $ appears in both semileptonic
and factorized hadronic decays, the factorization hypothesis can be tested 
by comparing these two processes. Eliminating the common matrix terms in these
decays, the following relation can be derived [2, 3, 57]:
$$ { { \Gamma(\bar B^{0} \rightarrow D^{*+} \pi^{-}) } \over 
 { { { d\Gamma} \over {dq^{2} } } 
 (\bar B^{0} \rightarrow D^{*+} l^{-} \bar {\nu_{l}} )|_{q^{2} = m_{\pi}^{2} 
} } }
=  6 \pi^{2} c_{1}^{2} f_{\pi}^{2} |V_{ud}|^{2}  $$
$$ = 1.22 \pm 0.15~ (theory), ~1.14 \pm 0.21~ ~(Expt.).  \eqno(38a) $$
Here, we require that the lepton -neutrino system has the same 
kinematic properties as does the pion in hadronic decay.
Similar relations can be obtained for $\bar B^{0} \rightarrow D^{*} \rho$
and $\bar B^{0} \rightarrow D^{*} a_{1}$ decays, 
$$ { {\Gamma(\bar B^{0} \rightarrow D^{*+} \rho^{-}) } \over 
  {{ { d\Gamma} \over {dq^{2} }} 
(\bar B^{0} \rightarrow D^{*+} l^{-} \bar {\nu_{l}})|_{q^{2} = m_{\rho}^{2}
 }} }
=  6 \pi^{2} c_{1}^{2} f_{\rho}^{2} |V_{ud}|^{2}  $$
$$ = 3.26 \pm 0.42~ (theory), ~2.80 \pm 0.69~ ~(Expt.), \eqno(38b) $$
$$ { {\Gamma(\bar B^{0} \rightarrow D^{*+} a_{1}^{-}) } \over 
{ { { d \Gamma} \over {dq^{2}} } 
( \bar B^{0} \rightarrow D^{*+} l^{-} \bar {\nu_{l}})|_{q^{2} = 
m_{a_{1}}^{2} }} }
 =  6 \pi^{2} c_{1}^{2} f_{a_{1}}^{2} |V_{ud}|^{2}  $$
$$ = 3.0 \pm 0.5 ~ (theory), ~3.6 \pm 0.9~ ~(Expt.). \eqno(38c) $$
Theory agrees well with experiment with in errors.
\subsubsection { Application of Factorization}
Having factorization tested, one may exploit this to extract information about
poorly measured semileptonic decays. For example, 
integrating over $q^{2}$-dependence and 
using experimental value $B( B^{-} \rightarrow D^{**0} \pi^{-}) = 
0.15 \pm 0.05$, one obtains [3]: 
$$B( B \rightarrow D^{**} l \nu ) = 0.48 \pm 0.16 \%~~ (1.00\pm 0.30 \pm 0.07 ~
~Expt.). \eqno(39) $$
Another application of relating hadronic mode with semileptonic decay is to 
determine $f_{D_{s}}$. For instance,  
$B(B^{0} \rightarrow D^{*+} D_{s}^{-}) = 0.93 \pm 0.25 \%$ gives [3]
$$f_{D_{s}} = 271 \pm 77 MeV. \eqno(40) $$
using $ B(D_{s} \rightarrow \phi \pi^{+}) = 3.7 \%.$ Similarly, one can obtain
$$f_{D_{s}^{*}} = 248 \pm 69 MeV. \eqno(41) $$
\subsubsection { Results from Heavy Quark Effective Theory}
Spin symmetry, appearing in the limit of heavy quark mass, combined with
factorization relates different decays [3]. For instance,
$$ {{ B(\bar B^{0} \rightarrow D^{+} \pi^{-}) } \over 
{ B(\bar B^{0} \rightarrow D^{*+} \pi^{-}) }} = 1.03 
~(1.11 \pm 0.22 \pm 0.08 ~Expt.)  \eqno(42) $$
$$ {{ B(\bar B^{0} \rightarrow D^{+} \rho^{-}) } \over 
{ B(\bar B^{0} \rightarrow D^{*+} \rho^{-}) }} = 0.89 
~(1.06 \pm 0.27 \pm 0.08 ~Expt.)  \eqno(43) $$
Using a combinations of HQET, factorization and data on semileptonic decay
$B \rightarrow D^{*} l \nu_{l}$, Mannel et al. [58] have obtained the following
predictions for 
$$ {{ B(\bar B^{0} \rightarrow D^{+} \rho^{-}) } \over 
{ B(\bar B^{0} \rightarrow D^{+} \pi^{-}) }} = 3.05, 2.52, 2.61  \eqno(44) $$
for three different parameterizations of the Isgur-Wise function.
Experimental value for this ratio is 
$$ {{ B(\bar B^{0} \rightarrow D^{+} \rho^{-}) } \over 
{ B(\bar B^{0} \rightarrow D^{+} \pi^{-}) }} = 2.7 
 \pm 0.6. ~(Expt.)  $$
Similarly predictions have also been made for 
$ B \rightarrow D D_{s} / D^{*} D_{s} / D^{*} D_{s}^{*} $ decays [3].
\subsubsection {Rare B-Decays}
Charmless decays involving $ b \rightarrow u $ transition, like 
 $ B \rightarrow \pi \pi / \pi \rho / K \pi $, are important to find 
$V_{ub}$, probe penguin contributions and to study CP-violation [3, 59]. 
Weak radiative B-meson decays present a very sensitive probe
of new physics, like Supersymmetry particle contributions. 
Precise measurements of exclusive radiative decays, like 
$B \rightarrow K^{*} \gamma$, would throw light on $V_{tq}$ elements [2, 3]. 
B-mesons have enough energy to create
p-wave mesons also. Branching rations of such decays have been estimated
using the ISGW model [60]. B mesons provide an unique opportunity to study
baryon-antibaryon decays of a meson. However, only a few upper limits are
available experimentally [1, 61]. There is now a considerable experimental 
evidence for $B-\bar B$ oscillations, which can be used to determine 
$V_{td}$ and $V_{ts}$ elements [2, 3]. 
\subsection {Weak Hadronic Decays of Baryons}
For heavy flavor baryon decays, data has only recently started coming in. 
Two-body decays of the baryons are of the following types:
$$B( {1/2}^{+}) \rightarrow B( {1/ 2 }^{+})/D(3/2^{+}) + 
P( 0^{-})/V(1^{-}). $$
Experimentally, branching ratios of almost all the Cabibbo enhanced
 $\Lambda^{+}_{c}\rightarrow B({1\over 2}^{+})+P(0^{-})$
decays have now been
measured [1, 44]. A recent CLEO measurement [62] of decay asymmetries of
$\Lambda^{+}_{c} \rightarrow \Lambda \pi^{+}/\Sigma^{+} \pi^{0} $,
give the following sets of PV and PC amplitudes (in units of
$G_{F} V_{ud}V_{cs} \times 10^{-2} GeV^{2})$:
$$ A(\Lambda^{+}_{c} \rightarrow \Lambda \pi^{+}) = -3.0^{+0.8}_{-1.2}  
\qquad \hbox{ or }   -4.3 ^{+0.8}_{-0.9},  $$
$$ B(\Lambda^{+}_{c} \rightarrow \Lambda \pi^{+})= +12.7^{+2.7}_{-2.5}  
\qquad \hbox { or }  +8.9^{+3.4}_{-2.4};   $$
$$ A(\Lambda^{+}_{c} \rightarrow \Sigma^{+} \pi^{0})= +1.3^{+0.9}_{-1.1} 
\qquad \hbox { or }  +5.4 ^{+0.9}_{-0.7}, $$
$$ B(\Lambda^{+}_{c} \rightarrow \Sigma^{+} \pi^{0})= -17.3^{+2.3}_{-2.9}  
\qquad \hbox { or }   -4.1 ^{+3.4}_{-3.0}.$$

Recently, CLEO-II experiment [63] has measured $ B(\Xi_{c}^{+} \rightarrow 
\Xi^{0} \pi^{+}) = 1.2 \pm 0.5 \pm 0.3 \%$.
This small data has already shown discrepancies with conventional
expectations. In the beginning, it was thought that like charm meson decays,
charm baryon decays may be dominated by the spectator quark process. 
This scheme allows only the emission of $\pi^{+}/\rho^{+}$ and $\bar K^{0}
/ \bar K^{*0}$ mesons. However, 
observation of certain decays like $\Lambda_{c}^{+} \rightarrow \Xi^{0} K^{+}/
\Sigma \pi, \Sigma \eta $ gives clear indication of the nonspectator 
contributions.
In fact, W-exchange quark diagram, suppressed in the meson decays due to 
the helicity arguments, can play a significant role due to the
appearance of spin 0 two-quark configuration in the baryon structure.
Due to the lack of a straightforward method to evaluate these terms, flavor
 symmetry [64] and model calculations [65] have been performed. 
So far no theoretical model could explain the experimental values.

Study of bottom baryon decays is just beginning 
to start its gear. So far, only exclusive weak hadronic decay 
$ \Lambda_{b} \rightarrow J/\psi + \Lambda $ has been measured. 
Recent CDF Collaboration experiment [66] gives 
$ B( \Lambda_{b} \rightarrow J/\psi + \Lambda ) = (3.7 \pm 1.7 \pm 0.4)
\times 10^{-4} $ which is consistent with theoretical expectation [67].
\section {Conclusions}
In the last several years, tremendous progress has been achieved 
in understanding the heavy flavor weak decays. We make the following 
observations:\\
\vskip 1mm 
1) Leptonic decays are the simplest to be treated theoretically, 
but have very small branching ratios. 
Since a direct determination of meson decay constants
 is highly desirable, particularly for $B-\bar B$ mixing, it is important 
to improve their measurements as larger data samples are accumulated.\\
\vskip 1mm 
2) Semileptonic decays are next in order of simplicity from theory side.
   Here all the strong interaction effects are expressed in terms of a few
   formfactors, which are reasonably obtained in theoretical calculations,
based on quark models, HQET, lattice simulation and QCD sum-rule approaches.
However, higher precision measurements are needed to find $V_{ub}$.\\
\vskip 1mm 
3) Weak hadronic decays experience large interference due to the strong
 interactions and pose serious problems for theory, particularly for the
charm hadrons. Though qualitative 
explanation can be obtained for these decays, discrepancies between theory
and experiment indicate the need of additional physics.  For instance,
final state interaction effects play significant role at  least in the
charm meson decays. Smearing effects due to the large width help to improve 
the agreement when a wide resonance, like $\rho$ is emitted in a decay.\\
\vskip 1mm 
4) Results from CLEO II have significantly modified our understanding of
weak hadronic B-decays. Data on their branching are now of sufficient quality
to perform nontrivial tests of factorization hypothesis. It seems to be 
consistent at the present level of experiment. 
Large sample of B-decay data will be obtained in next few years which
will present more accurate tests for the factorization scheme.\\
\vskip 1mm 
5) The ratio $a_{2}/a_{1}$ is demanded to be positive for bottom meson decays 
  in contrast to what is found in the charm meson decays. 
This has opened the issue of nonfactorizable terms for the weak hadronic 
decays.
It is now clear that significant nonfactorizable contributions are there in 
the weak hadronic decays of charm mesons. For bottom sector, 
a quantitative estimate of their size require precise measurements of their
decays. Study of rare decays, like radiative decays and charmless B-decays,
 has a good potential to throw new lights on our
understanding of the penguin terms and CP violation.\\
\vskip 1mm
6) Weak hadronic decays of charm baryon have recently come under active 
experimental investigation, though search for bottom baryon  decays is merely
begun. These decays are difficult to treat theoretically. Observed data
for $\Lambda_{c}$ decays clearly demand significant W-exchange contributions.
More data on baryon decays, which will be accumulated in the near future, 
is expected to confront theory with new challenges.
\vskip 1mm
{\bf Acknowledgments}\\
Financial assistance from the Department of Science \& Technology, New Delhi
(India) is thankfully acknowledged.


\newpage
\begin{thebibliography}{99}
\bibitem[1]{} R.M. Barnet et al.,
Particle Data Group, Phys. Rev. D {\bf 54}, 31 (1996). 
\bibitem[2]{} "heavy Flavors", A.J. Buras and H. Lindner (eds.),
World Sci. Singapore, (1992); "B-Decays", S. Stone (ed.), 
World Sci. Singapore, (1994). 
\bibitem[3]{} T.E. Browder and K. Honscheid, "B Mesons", UH 511-816-95, 
appeared in Prog. Nucl. Part. Phys. {\bf 35} (1996).
\bibitem[4]{} J.D. Richman and P.R. Burchat, Rev. Mod. Phys. {\bf 67} 893, 
(1995).  
\bibitem[5]{} G. Martinelli, "Theoretical Review of B Physics", Rome prep-
1155/96.
\bibitem[6]{} M. Wirbel, Prog. Nucl. Part. Phys. {\bf 21}, 333 (1988).
\bibitem[7]{}  A.F. Falk, M.B. Wise and I. Dunietz, Phys. Rev. D {\bf 51}, 
1183 (1995); I.I. Bigi, UND-HEP-96-BIG06,(1996), hep-ph/9612293.
\bibitem[8]{} S.N. Ganguli, "Physics for LEP 1", and 
A. Gurtu, "Physics for LEP 2", presented at this symposium.
\bibitem[9]{} V. Jain, "Recent Results from CLEO", presented at this symposium.
\bibitem[10]{} T. Aziz, "Heavy Flavor Physics", presented at this symposium.
\bibitem[11]{} P. Quintas et al., " The Standard Model and Beyond", Fermilab-
FN-640 (1996); S.D. Rindani, "New Physics at $e^{+} e^{-}$ colliders", 
presented at this symposium.
\bibitem[12]{} P. Colangelo and F.De. Fazio, Phys. Lett. B {\bf 87}, 371 
(1996).  
\bibitem[13]{} G. Altarelli, N. Cabibbo, L. Maiani, Phys. Lett. B {\bf 382}, 
409 (1996); Author thanks A. Kundu for raising this point during the 
discussion.
\bibitem[14] {} M. Wirbel, B. Stech and M. Bauer, Z. Phys. C {\bf 29}, 637 
(1985); M. Bauer, B. Stech and M. Wirbel, {\it ibid } {\bf 34}, 103 (1987). 
\bibitem[15]{} M. Bauer and M. Wirbel, Z. Phys. C {\bf 42}, 671 (1989). 
\bibitem[16]{} W. Jaus, Phys. Rev. D {\bf 41}, 3394 (1990); D. Melikov, 
{\it ibid } {\bf 53}, 2460  (1996), Phys. Lett. B {\bf 380}, 363 (1996);
N. Barik and P.C. Dash, Phys. Rev. D {\bf 53}, 1366 (1996); 
R.N. Faustov et al., {\it ibid } {\bf 53}, 1391 (1996).   
\bibitem[17] {} N. Isgur, D.
Scora, B. Grinstein and M. Wise, Phys. Rev. D {\bf 39}, 799 (1989).
\bibitem[18]{}  D. Scora and N. Isgur, Phys. Rev. D {\bf 52}, 2783 (1995). 
\bibitem[19]{} T. Altomari and L. Wolfenstein, Phys. Rev. D {\bf 37}, 681 
(1988). 
\bibitem[20]{} G. Altarelli et al., Nucl. Phys. B {\bf 208}, 365 (1982);
J.G. K\"orner and G.A. Schuler, Phys. Lett. B {\bf 226}, 185 (1989); 
Z. Phys. C {\bf 46}, 93 (1990); F.J. Gilman and R.L. Singleton Jr.,
Phys. Rev. D {\bf 41}, 93 (1990); D.K. Choudhury et al., Pramana 
{bf 44}, 519 (1995); D.K. Choudhury and P. Das, {\it ibid} 
{bf 46}, 349 (1996).    
\bibitem[21]{} M. Neubert, Phys. Rep. {\bf 245}, 259 (1994); 
F.E. Close and A. Wambach, Nucl. Phys. B {\bf412}, 169 (1994).
\bibitem[22]{} C.W. Bernard et al., Phys. Rev. D {\bf 43},2140 (1991),
{\it ibid } {\bf45}, 869 (1992), {\it ibid } {\bf 47}, 998 (1993);   
 A.  Abada et al., Nucl. Phys. B {\bf 376}, 172 (1992),
{\it ibid } {\bf 416},675 (1994); C.R. Allton et al., Phys. Lett. B {\bf 326}, 
295 (1994);  J.M. Flynn, "Developments in Lattice QCD", Southampton Univ. 
Report SHEP-96-33 (1996).
\bibitem[23]{} P. Ball, V.M. Braun and H.C. Dosch, Phys. Rev. D {\bf 44}, 
3567 (1991); B. Blok and M. Shifman, {\it ibid } {\bf 47}, 2949 (1993); 
 S. Narison, Phys. Lett. B {\bf 325}, 197 (1994). 
\bibitem[24]{} R.E. Marshak, Riazuddin and C.P. Ryan, "Theory of Weak 
Interactions in Particle Physics", Wiley, N.Y. (1969).
\bibitem[25]{} D. Green, "Lectures in Particle Physics", 
World Sci. Singapore, (1994). 
\bibitem[26]{} J.G. Adler et al., Phys. Rev. Lett. {\bf 60}, 1375 (1988). 
\bibitem[27]{} M. Suzuki, Phys. Rev. D {\bf 54}, 319 (1994). 
\bibitem[28]{} C.T. Sachrajda, "Exclusive Decays of Beauty Hadrons", 
CERN-TH/96-257 (1996).
\bibitem[29]{} E653 Collaboration: K. Kodama et al., hep-ex/9606017 (1996). 
\bibitem[30]{} CLEO Collaboration: D. Gibaut et al., CLEO-conf-95-22 (1995).
\bibitem[31]{}  H. Schr\"oder, "$B \bar B$ mixing", in ref [2].
\bibitem[32]{} Y. Grossman and M.P. Worah, SLAC-PUB-7351, (1996), hep-ph/
9612269.
\bibitem[33]{} A. Ali and C. Greub, Z. Phys. C {\bf 60}, 433 (1993). 
\bibitem[34]{} R.J. Morrison and J.D. Richman, Phys. Rev. D {\bf 50}, 
1565 (1994).  
\bibitem[35]{} J.D. Richman, Phys. Rev. D {\bf 54}, 482 (1996). 
\bibitem[36]{} CLEO Collaboration: J.P. Alexander et al., 
"First Measurements of $B \rightarrow \pi l \nu$ and $ B \rightarrow 
\rho(\omega) l \nu$ Branching", CLNS 96-1419, CLEO -96-6 (1996).
\bibitem[37]{} J.G. K\"orner, M. Kr\"amer and D. Pirjol, Prog. Part. Nucl.
Phys. {\bf 33}, 787 (1994).
\bibitem[38]{} M.J. Savage and M.B. Wise, Phys. Rev. D {\bf 39}, 3346 (1989);
Y. Kohara, Phys. Lett. B {\bf 228}, 523 (1989); 
S.P. Rosen, {\it ibid } {\bf 228}, 525 (1989);
Phys. Rev. D {\bf 41}, 303 (1990);    
R.C. Verma and A.N. Kamal, {\it ibid } {\bf 43}, 829 (1990);  
A.N. Kamal, R.C. Verma and N. Sinha, {\it ibid } {\bf 43}, 843 (1990).
S.M. Sheikholeslami and R.C. Verma, Int. J. Mod. Phys. A {\bf 7} 3691 (1992);
A.C. Katoch and R.C. Verma, Ind. J. Pure. App. Phys. {bf 31}, 216 (1993).
\bibitem[39]{} F. Buccella et al., {\it ibid } {\bf 51}, 3478 (1995);
L. Hinchliffe  and T.A. Kaeding, {\it ibid } {\bf 54}, 914 
(1996).
\bibitem[40]{}  L.L Chau and H. Y. Cheng, Phys. Rev. D {\bf 42 }, 
1837 (1990), {\it ibid } {\bf 43} 2176 (1991); 
Y. Kohara, {\it ibid }{\bf 44}, 2799 (1991); Zhi-Zhong Xing, "Remarks 
on Quark Diagrams Decays in Two-body Nonleptonic B-Meson Decays", Univ.
M\"unchen prep.- LMU-13/94.
\bibitem[41]{} A.J. Buras, J.M. Gerard and R. R\"uckl, 
Nucl. Phys. B {\bf 268}, 16 (1986).
\bibitem[42]{} J.D. Bjorken, Nucl. Phys. (Proc. Supp.) {\bf 11}, 325 (1989);
 A.N. Kamal, Q.P. Xu and A. Czarnecki, Phys. Rev. D {\bf 49}, 1330 (1994). 
\bibitem[43]{} D.M. Asner et al., Phys. Rev. D {\bf 53}, 1039 (1996). 
\bibitem[44]{} CLEO Collaboration: J.P. Alexander et al., Phys. Rev. D 
{\bf 53}, R1013 (1996). 
\bibitem[45] {} N.G. Deshpande, M. Gronau and D. Sutherland, Phys. Lett. 
B {\bf 90}, 431 (1980); H.Y. Cheng, Z. Phys. C. {\bf 32}, 237 (1986), 
J.M. Soares, Phys. Rev. D {\bf 51}, 3518 (1995);
A.N. Kamal et al., Phys. Rev. D {\bf 53 }, 2506 (1996).            
\bibitem[46] {} R.C. Verma, Phys. Lett. B {\bf 365}, 377 (1996);
K.K. Sharma, A.C. Katoch and R.C. Verma, Z. Phys. C,  (1997) 
in press. 
\bibitem[47] {} R. C. Verma, A. N. Kamal and M. P. Khanna,
Z. Phys. C. {\bf 65}, 255 (1995); R. C. Verma, in  Proc.  Lake Louise
Winter Institute on `Quarks and Colliders', A. Astbury et al., (eds),
 World Sci. Singapore (1996).
\bibitem[48]{} A.N. Kamal, Int. J. Mod. Phys. A {\bf 7}, 3515 (1992);
 A.N. Kamal, N. Sinha and R. Sinha, Z. Phys. C {\bf 41}, 207 
(1988); A. N. Kamal and T. N. Pham, Phys. Rev. D, {\bf 50}, R1832 
(1994); X.Q. Li and B.S. Zou, " Significance of Single Pion Exchange:
Inelastic FSI for D $\rightarrow $ PV" RAL-TR-96-079 (1996);
F. Buccella et al., "Charm Nonleptonic Decays and Final State Interactions", 
Napoli- DSF-T-2196, hep-ph/9601343 (1996).
\bibitem[49]{} S. Malvezzi, "Analysis of Substructure in Charm Decays", 
Frascati Series {\bf XXX} (1997) to appear.
\bibitem[50]{} T. Uppal and R.C. Verma, Phys. Rev. D {\bf 46}, 2982 (1992). 
\bibitem[51]{} T. Uppal and R.C. Verma, Z. Phys. C {\bf 56}, 273 (1992). 
\bibitem[52] {} M. Gourdin, A. N. Kamal, Y. Y. Keum and X. Y. Pham, Phys.
Letts. B {\bf 333}, 507 (1994); CLEO Collaboration: M.S. Alam {\it et al.}, 
Phys. Rev. D {\bf 50}, 43 (1994); D. G.  Cassel, in  Proc.  Lake Louise
Winter Institute on `Quarks and Colliders', A. Astbury et al., (eds),
 World Sci. Singapore (1996).
\bibitem[53]{} B. Blok and M. Shifman, Nucl. Phys. B {\bf 399}, 441 (1993);
{\it ibid } {\bf 389}, 534 (1993); A. Khodjamirian and R. R\"uckl, in "QCD 94",
 Proc. Int. Conf. 
Montpellier, France (1994) ed. S. Narison, Nucl. Phys. B Proc. Supp. {\bf38},   
396 (1990).
\bibitem[54]{} R.C. Verma Z. Phys. C {\bf 69}, 253 (1996); 
 A.C. Katoch, K.K. Sharma and R.C. Verma, J. Phys. G (1997) to appear.
\bibitem[55]{} A.N. Kamal and R.C. Verma, Phys. Rev. D {\bf 45}, 982 (1992);
X.Y. Pham and X.C. Vu, {\it ibid } {\bf 46}, 261 (1992); 
F. Buccella et al., Z. Phys. C {\bf 55}, 982 (1992); 
A.C. Katoch and R.C. Verma, Phys. Rev. D {\bf 49}, 1645 (1994);
Z. Phys. C {\bf 62}, 173 (1994); J. Phys. G {\bf 21}, 525 (1995);   
A.N. Kamal and  Q.P. Xu, Phys. Rev. D {\bf 49}, 1526 (1994); 
\bibitem[56]{} T.E. Browder, "Hadronic Decays and Lifetimes of B and D Mesons"
 Univ. Hawaii-UH-511-857-96 (1996).
\bibitem[57]{} D. Bortoletto and S. Stone, Phys. Rev. Lett. {\bf 65}, 2951
 (1990); P. Colangelo, G. Nardulli and N. Paver, Phys. Lett. B
 {\bf 293}, 207 (1992); P. Colangelo, F. De. Fazio and G. Nardulli,
Phys. Lett. B {\bf 303}, 152 (1993). 
\bibitem[58]{} T. Mannel et al., Phys. Lett. B {\bf 259}, 359 (1991). 
\bibitem[59]{} R. Enomoto and M. Tanabashi, "Direct CP Violation of 
B Mesons via $ \rho - \omega $ interference", Fermilab- PUB-96/130-T;
S.Y. Grossman and M.P. Worah, SLAC - PUB- 7351, hep-ph/9612269.
\bibitem[60]{} A.C. Katoch and R.C. Verma, Phys. Rev. D {\bf 52}, 1717 (1995);
Int. J. Mod. Phys. A {\bf 41}, 129 (1996); J. Phys. G {\bf 22}, 1765 (1996);
G. Lopez Castro and J.H. Munoz, Phys. Rev. D (1997) to appear;    
\bibitem[61]{} CLEO Collaboration: X. Fu et al., "Observation of Exclusive 
B Decays to Final State containing a Charm Baryon", CLNS 96/1397, CLEO -96-6 
(1996).
\bibitem[62]{} CLEO Collaboration: M. Bishai, {\it et al.}, Phys. Lett. B
{\bf 350}, 256 (1995).
\bibitem[63]{} CLEO Collaboration: K.W. Edwards, {\it et al.}, Phys. Lett. B
{\bf 373}, 261 (1996).
\bibitem[64]{} M.J. Savage and R.P. Singer, Phys. Rev. D {\bf 42}, 1527 (1990);
S. M. Sheikholeslami, et al., {\it ibid } {\bf 43}, 
170 (1991);  J.G. K\"orner and M. Kr\"amer, Z. Phys. C {\bf 55}, 659 (1992);
R.C. Verma and M.P. Khanna, {\it ibid } {\bf 53}, 3723 (1996). 
\bibitem [65]{} S. Pakvasa, S. F. Tuan, and S. P.  Rosen, Phys. Rev. D 
{\bf 42}, 3746 (1990);
 G. Turan and J. O. Eeg, Z. Phys. C {\bf 51 }, 599 (1991);
 R. E.  Karlsen and M. D. Scadron, Euro Phys. Lett. {\bf 14 }, 319 (1991);
  J. G. K\"orner and H. W. Siebert, Ann. Rev. Nucl. Part. Sci. {\bf 41}, 511 
(1991);  G. Kaur and M. P. Khanna, Phys. Rev. D {\bf 44}, 182 (1991), 
 {\it ibid } {\bf 45}, 3024 (1992);
  Q. P. Xu and A. N. Kamal, {\it ibid } {\bf 46}, 270 and 3836 (1992);
 H. Y. Cheng and B. Tseng, {\it ibid } {\bf 46}, 1042 (1992); 
 {\it ibid } {\bf 48}, 4188 (1993). 
 H. Y. Cheng {\it et al.}, {\it ibid } {\bf 46 }, 5060 (1992);
T.M. Yan et al., Phys. Rev. D {\bf 46}, 1148 (1992); 
 P. Zencykowski, {\it ibid } {\bf 50 }, 402 (1994).
  T. Uppal, R. C. Verma, and M. P. Khanna, {\it ibid } {\bf 49}, 3417 (1994).
\bibitem[66]{} CDF Collaboration: F. Abe et al., Fermilab-Pub - 96/270-E. 
\bibitem[67] {}  H.Y. Cheng, "Nonleptonic Weak Decays of Bottom Baryons",
IP-ASTP -06-96 (1996).
\end{thebibliography}
\end{document}